\documentclass[english,prb,preprint,superscriptaddress]{revtex4}

\usepackage{times}
\usepackage[T1]{fontenc}
\usepackage[latin1]{inputenc}
\usepackage{amsmath}
\usepackage{amssymb}	
\usepackage{MnSymbol}
\usepackage{xfrac}
\usepackage{float}
\usepackage{color}
\usepackage{graphicx}
\usepackage{xfrac}
\makeatletter
\usepackage{babel}
\makeatother

\begin{document}

\title{Native defects in ultra-high vacuum grown graphene islands on Cu(111)}

\author{S. M. Hollen}\email{hollen.8@osu.edu}
\author{S. J. Tjung}
\author{K. R. Mattioli}
\author{G. A. Gambrel}
\author{N. M. Santagata}	
\author{E. Johnston-Halperin}
\author{J. A. Gupta}
\affiliation{The Ohio State University}

\begin{abstract} 
We present a scanning tunneling microscopy (STM) study of native defects in graphene islands grown by ultra-high vacuum (UHV) decomposition of ethylene on Cu(111).  We characterize these defects through a survey of their apparent heights, atomic-resolution imaging, and detailed tunneling spectroscopy.  Bright defects that occur only in graphene regions are identified as C site point defects in the graphene lattice and are most likely single C vacancies. Dark defect types are observed in both graphene and Cu regions, and are likely point defects in the Cu surface.  We also present data showing the importance of bias and tip termination to the appearance of the defects in STM images and the ability to achieve atomic resolution. Finally, we present tunneling spectroscopy measurements probing the influence of point defects on the local electronic landscape of graphene islands.  
 \end{abstract}

\maketitle

\section{Introduction}
For logic applications looking to take advantage of graphene's high mobility charge carriers, much recent work has focused on overcoming two main challenges: incorporation of intrinsic dopants and introduction of a band gap in the electronic structure. Both challenges can be approached by manipulating graphene's crystal structure. Compared to bulk, two-dimensional materials offer a unique advantage since the electronic structure of the entire material can be modified through functionalization of its surface.  This functionalization can proceed from adatom binding, the creation of lattice defects, or the substitution of atoms at C sites.\cite{Boukhvalov2009,Wang2013a}  Hydrogen functionalization of graphene (to create graphane) shows evidence of opening a band gap,\cite{Balog2010, Elias2009} as was predicted by density functional theory calculations.\cite{Sofo2007}   Sparse single vacancies are shown to be effective sites for introducing dopants while maintaining reasonable transport properties.\cite{Wang2012,Pantelides2012,Bangert2015,Vicarelli2015,Peres2006,Lherbier2008}  Di-vacancies are also considered attractive sites for substitutional doping due to the size and band structure effects of certain potential dopants (such as Al, P, As, Ga, In, Pt, and Co), but experimental work is still limited.\cite{Tsetseris2014,Ugeda2012,Wang2012}
Most experimental work so far has involved boron and nitrogen substitutions at single C sites.\cite{Wei2009,Guo2010,Zhao2011a,Zhao2013,Tison2015}  These are either incorporated into the lattice during chemical vapor deposition growth\cite{Wei2009,Zhao2013} or afterward by ion irradiation and annealing\cite{Guo2010} and provide p- and n-type doping, respectively.  A common challenge in functionalizing graphene through defect engineering is in the defects' relatively high formation energy and their tendency to cluster,\cite{Pantelides2012,Kotakoski2011} which further motivates study in defect creation and control.

Defects in graphene are also of fundamental interest. Single vacancies were predicted to have magnetic moments,\cite{Yazyev2007} and there is now strong evidence for such.\cite{Ugeda2010,Nair2012}  Point defect engineering may also enable control of plasmons on sub-wavelength scales.\cite{Zhou2012}  From a structural point of view, graphene and other 2D crystals provide new opportunities to directly correlate defects and macroscopic properties such as transport. For example, Qi and coworkers\cite{Qi2015} used TEM and transport measurements to directly correlate defect healing at the edges of graphene nanoribbons with increased conductivity. Prior TEM studies imaged the formation and mobility of graphene defects in real-time,\cite{Meyer2008} studied the insertion of interstitials,\cite{Lehtinen2015} and probed doping via vacancies created with ion bombardment.\cite{Wang2012}  Because of their relatively high formation energy,\cite{Banhart2011} graphene and graphite defect studies are typically limited to defects that are created by ion bombardment and irradiation.\cite{Ziatdinov2014,Wang2012,Meyer2008,Liu2015}  As a stochastic process, the defect density and clustering produced by ion bombardment provide limited control. Native defects created during growth may allow for better control, but graphene grown in near-equilibrium conditions (such as CVD) is unlikely to exhibit native lattice defects, except at grain boundaries where independently nucleated growth regions meet.  In one study, defective graphene was achieved at intermediate phases during growth in UHV; the vacancy concentration could then be controlled by healing of vacancies during subsequent thermal annealing. \cite{Niu2013}

Here we report a scanning tunneling microscopy study of native graphene defects that form during ultrahigh vacuum graphene growth on Cu(111).  The growth proceeds by ethylene decomposition on a hot Cu(111) crystal in short (5-10 minute) cycles.\cite{Gao2010,Zhao2011,Jeon2011}  The UHV-grown graphene is then studied without exposure to air by {\em in situ} transfer into the STM chamber. We conducted a survey of over 100 point defects on 15 different graphene regions from 3 similar graphene growths.  Atomic resolution imaging helps us identify bright defects as graphene lattice point defects, likely single vacancies. We present evidence showing that dark defects are most likely point defects in the underlying Cu lattice. The imaging contrast and resolution are discussed with respect to bias voltage and the STM tip's atomic-scale termination. We also present dI/dV spectroscopy probing the local density of states (LDOS) over point defects, which reveals a relatively limited contribution to the local electronic landscape.

\section{Methods}
Reproducible, epitaxial growth of graphene islands on Cu(111) is achieved by thermal decomposition of ethylene in ultra-high vacuum (UHV) conditions. This method produces pristine graphene that can be studied by STM without exposure to air. We first prepare a clean Cu(111) substrate by repeated cycles of Ar$^+$ sputtering and annealing at 600$^\circ$C in 5-minute intervals to remove surface contamination. Graphene is then grown by introducing ethylene gas at a pressure of 2 x 10$^{-5}$ mbar into the preparation chamber while cycling the sample temperature from room temperature to 950$^\circ$C 2-4 times. Each temperature cycle is performed in 5 minutes and we pump the ethylene out of the chamber when the sample temperature is at $\sim$500$^\circ$C during the cooldown of the last cycle. This procedure leverages the enhanced sticking of ethylene on Cu at room temperature, while promoting the the thermal decomposition and self-assembly of graphene at high temperature. The sample is immediately cooled to 100 K and transferred into the STM chamber where it further cools to 5 K in $\sim$6 hours.  
 
The resulting graphene islands on Cu(111) are studied with a CreaTec STM at 5 K. Using STM, we can probe topographic and spectroscopic information at an atomic scale.  Images presented here were obtained by measuring small changes in height of the STM tip to maintain constant current while scanning across the sample surface. A cut PtIr tip was used for all of these measurements. For spectroscopy measurements, the tunneling current was recorded while sweeping the bias voltage at a tip height fixed by the initial tunneling conditions. By adding a small modulation of 5-10 mV$_\mathrm{pp}$ at $\sim$1 kHz to the bias voltage, a lock-in amplifier simultaneously measures the differential conductance, dI/dV, proportional to the convolved tip and sample local density of states (LDOS). Spatial maps of dI/dV at a specific voltage can be collected simultaneously with topographic images by recording the lock-in signal pixel-by-pixel during scanning.

\section{Results}
Five representative examples of graphene islands exhibiting native defects are displayed in Figure 1. Graphene is most often observed near step edges, and can grow continuously over terraces, as shown by the continuous Moire pattern over three terraces of Fig. 1b.  Islands often grow with a regular, hexagonal shape as shown in Fig. 1c. A range of island sizes was observed, with an average area of 940 nm$^2$. The point defects in the graphene regions can be classified as either bright or dark by their contrast in STM images (c.f., yellow and blue arrows in Fig. 1a respectively). We surveyed  93 bright and 29 dark defects on 15 graphene islands from three similar, but independent growth runs. Additional features are sometimes seen, such as dark clusters of defects (Fig 1b, red arrow) and irregular lines (Fig 1e, green arrow), which are found to be grain boundaries between graphene regions when imaged with atomic resolution.   

On graphene, we find an average of one dark defect per 450 nm$^2$. A much higher density of dark point defects is seen on the Cu (e.g., Fig. 1b), and are commonly attributed to surface contaminants such as CO\cite{Bartels1999} or O.\cite{Wiame2007} Auger spectroscopy of the sample after growth (not shown) indicates traces of S and O which may segregate from the bulk to the surface under the growth conditions.  We also expect an amount of unreacted carbon from the ethylene decomposition. The dark point defects observed in the graphene regions have similar apparent depths and lateral extent as those observed on the neighboring Cu(111) substrate. They are also consistent with those observed for graphene on Cu(111) by Mart\`{i}nez-Galera et al.,\cite{Martinez-Galera2011} which were attributed to defects in the underlying Cu surface. We will revisit this assignment below when we discuss atomic resolution images. 

The more common bright defects are only found on the graphene regions with a density of approximately one defect per 150 nm$^2$, and will be the primary focus of this report. STM does not have elemental specificity a priori, so indirect measurements are needed to gain further insight into the origins of the defect. For example, we carefully compared the apparent heights of the defects in STM images. The apparent heights of the bright defects vary from 8 pm to 21 pm under identical tunneling conditions (1 V, 0.2 nA), though they do not separate into clearly distinct groups. We attribute this range of measured apparent height to slight differences in tip termination, which become more important for small apparent heights like those measured here. To control for tip-dependent variations, we compare apparent heights of defects within the same image and imaged with the same tip termination. Figure 2a shows an example of a small hexagonal graphene island with 11 bright defects and 3 dark defects.  This island is one of the most defective graphene regions we observed. Representative line profiles over a bright and dark defect are shown in Figs. 2b and 2c, respectively. The apparent heights of the 11 bright defects span a range with an average and standard deviation of 17 $\pm$ 3 pm. This is a much smaller range than in our larger survey, and suggests that there is only a single origin for the bright defects we observe. 

To gain further insight into the atomic origins of the defects, we compared STM images under varying bias voltages (e.g., 1V, 0.05 V, and -1 V in Figures 2d-f). The bright defects in Figures 2d and 2f are similarly imaged as circularly symmetric protrusions, though the apparent height changes from 21 pm to 26 pm respectively.  At low bias (Fig. 2e), the bright defects only have an apparent height of $\sim$13 pm, and are largely obscured by scattering of Cu(111) surface state electrons. These electrons comprise a nearly free 2D electron gas (2DEG) at the surface.\cite{Crommie1993}  Recently, scattering of these surface state electrons was used to infer the charge state of adatoms on ultrathin films of NaCl grown on Cu(111).\cite{Repp2004} We observe no additional scattering of the surface state electrons due to the bright defects, which suggests that these defects are confined to the graphene layer, and have little or no effective charge. In contrast, the dark defects remain clearly visible in STM images at low bias, and we do observe circularly-symmetric scattering patterns of the surface state electrons. 

To probe the graphene lattice in the vicinity of the defects, we first note that atomic resolution depends strongly on the termination species of the STM tip, which is not known under most circumstances. Although the tip is PtIr in the bulk, Cu or other adsorbates can terminate the tip due to the repeated contact with the surface necessary to optimize imaging. The tip termination can be changed intentionally by contacting the sample surface and picking up surface atom(s), or by deliberately approaching e.g., adsorbed molecules such as CO under appropriate conditions.\cite{Hahn2001} Molecule-terminated tips often provide sharper resolution than metal-terminated tips due to introduction of additional chemical contrast. One of our motivations for using only sub-monolayer coverages of graphene, is that we have nearby clean Cu(111) that can be examined at the same time. In particular, we can look for anomalous features in tunneling spectroscopy such as prominent peaks that are often attributed to e.g., molecular orbitals. Furthermore, our experience suggests that atomic resolution on the close-packed Cu(111) surface is not achieved with metal-terminated tips under typical tunneling conditions; thus the appearance of the Cu lattice in STM imaging is a proxy for some kind of molecule-terminated tip. Because the tip termination often changes during the course of the experiments, we carefully look for changes in position or contrast during repeated imaging.  

STM images of an area imaged under the same tunneling conditions, but with three different tip terminations are shown in Fig. 3a-c. Tunneling spectroscopy and images on Cu suggest that Tip A was terminated with a metal atom (probably Cu), Tip B with an unknown molecule, and Tip C with a CO molecule. With Tip A, which is most common, the two defects in the image appear as bright, circularly symmetric protrusions. The same defects appear with atomic resolution when imaged with the molecule-terminated Tips B and C. For both tip terminations, the honeycomb lattice of graphene is clearly resolved and the defects are imaged as protrusions with trigonal symmetry.  

Additional substructure is shown in Figures 3d-e, which are STM images of the same defects with Tip B, but under different tunneling conditions (lower current). Since the tunneling conditions control both the tip-surface separation (via the current feedback loop) and the energy of tunneling electrons (via the applied bias), the resolution resulting from a molecule-terminated tip can depend strongly on these parameters. With this tip, we were able to tune the tunneling parameters to achieve highly-resolved images showing the surrounding graphene lattice and the internal structure of the bright defect. The graphene honeycomb lattice is clearly resolved around the defects, where the centers of the carbon rings image as protrusions.  By overlaying the graphene lattice on the images, we find that the defect is centered on a single carbon site, and has perfect trigonal symmetry extending from it.  Fig. 3e shows that the two defects, which are oriented in the same direction with respect to the graphene lattice, occur on the same sublattice site (blue dots).  Different orientations of the defects are also observed.  Figure 3f shows an atomic resolution image of a different graphene region imaged with a different tip (Tip D), that appears similar to Tip C.  Three triangular defects appear in Fig. 3f.  Two form upward-pointing triangles and the third forms a downward-pointing triangle.  Here, the honeycomb centers image as depressions, and we again overlay the graphene lattice to determine that the two defect orientations correspond to opposite sublattice sites (lower insets).  

Overall, the defect symmetry, orientation, and sublattice assignments agree with theoretical expectations for a single vacancy.\cite{Yazyev2007,Amara2007}  The three-fold scattering pattern from the defects arises because of suppressed intervalley scattering in graphene and reflects the symmetry of the sublattice.  Thus one would expect that defects on different sublattices would be rotated by 60$^\circ$, as observed.  We compare our images of single defects to simulated STM images of defects calculated using density functional theory (DFT)\cite{Yazyev2007} and a combined tight-binding approximation/DFT method\cite{Amara2007} for a freestanding graphene sheet.  Perfect trigonal symmetry is expected for an unreconstructed vacancy.\cite{Amara2007}  However, a reconstructed form of the vacancy, where the C-C interatomic distance across the vacancy shortens to reduce its overall energy, is expected.  Simulated STM images of the reconstructed vacancy show bright triangular features that lack three-fold symmetry,\cite{Amara2007,Yazyev2007} unlike what we observe. This suggests that the underlying Cu substrate stabilizes the symmetric, unreconstructed form of the defect. The alignment of the triangular features with respect to the graphene lattice imaged at low voltage is consistent with simulated images of the reconstructed vacancy reported by Yazyev and coworkers.\cite{Yazyev2007} However, we note that the region of Fig. 3e imaged with the same tip but at 1V (not shown) shows instead an orientation rotated by 60$^\circ$, consistent with the simulations of Amara and coworkers.\cite{Amara2007}  This difference highlights the importance of the bias voltage in interpreting experimental and simulated STM images. 

We note that nitrogen is a common residual gas in UHV, and one could assign the defects we observe to substitutional nitrogen instead. In fact, STM images of nitrogen\cite{Zhao2011a} substitutions also show the expected trigonal symmetry.\cite{Brito2012,Zheng2010} However, the experiments and DFT simulations of these substitutions show image contrast with symmetric bright spots at the three nearest-neighbor sites to the N atom.\cite{Zhao2011a} This difference and their larger apparent heights ($\sim$80 pm\cite{Zhao2011a}) makes them clearly distinguishable from single vacancies and our observations.  In our system, the most likely alternative to a single vacancy is a hydrogenated vacancy, which is also predicted to be perfectly three-fold symmetric.\cite{Yazyev2007}  However, in recent STM experiments and DFT simulations of Ar$^+$ created hydrogenated vacancies on graphite, several different configurations of hydrogen are shown with varying image contrast,\cite{Ziatdinov2014} where we see remarkable consistency in the appearance of the bright, triangular defects. These considerations suggest that the bright defects we observe are indeed single vacancies. 

Briefly returning to the dark defects, the white arrows in Fig. 3f indicate that these defects do not disturb the graphene lattice, which is continuous over the defects. We also notice that many of these types of defects appear to be swept to the edges of graphene islands during growth (see Ref. \cite{Hollen2015} and Fig. 1a, b, and c). These observations, along with the similarity between these defects and those on the bare Cu areas, and the scattering of the Cu surface state electrons (Fig. 2e, 3f) are strong evidence that these are defects in the underlying Cu lattice.\cite{Martinez-Galera2011} The larger dark area in the center of Fig. 3f appears to be quite different, and may be a defect cluster in the graphene. The upper inset of the figure shows this region with a Laplacian filter applied to emphasize the surrounding graphene lattice, which is clearly deformed at the edges of this feature. 

Lastly, we performed tunneling spectroscopy to probe how the defects influence the electronic structure of the graphene (Fig. 4). Point spectra taken on the centers of the dark and bright defects in Fig. 4a are compared to pristine graphene in Fig. 4b. The primary feature in these spectra is the band edge of the Shockley surface state in the underlying Cu(111), which is shifted to -0.31 V for graphene-covered Cu.\cite{Hollen2015} There are slight differences in the sharpness and magnitude of the band edge in dI/dV between the spectra on and off the defects. These differences are emphasized in a spatial map of the dI/dV signal at -0.31 V (Fig. 4c).  The bright defects show a greatly enhanced LDOS that extends symmetrically around the defect with a FWHM of $\sim$3 nm. The dark defects have a smaller enhancement of LDOS at their center, then a diminished LDOS in a ring shape surrounding the defect with a diameter of $\sim$1.2 nm. Ring-like structures in dI/dV maps like these have previously been attributed to ionization rings of charged defects, such as Co on graphene \cite{Brar2011} and Mn dopants in GaAs.\cite{Lee2011}  Additional studies, such as a bias dependence of the ring diameter, are needed to test whether this ring structure is due to ionization, which would be surprising given that the defects appear to be strongly coupled to the metal substrate. Figures 4d-e show color maps of dI/dV spectra, taken point-by-point by moving the tip along the lines indicated in Fig. 4a. These data indicate that the enhanced band edge LDOS extends only $\sim$1nm from both types of defects.There are no other prominent defect-specific electronic states within the range probed here.

Tunneling spectroscopy of defects in small, regularly shaped islands is further complicated by quantum confinement of the Shockley 2DEG. Figure 5 shows one such island, with a step in the underlying Cu running through the center. This island contains three bright defects, clearly resolved in Fig. 5b under imaging at 1V, but almost indiscernible in STM imaging at 0.05 V (Fig. 5c). Tunneling spectra taken in a line over the upper bright defect in Fig. 5b again shows the Shockley band edge at -0.31 V as the primary feature common to all the spectra (Fig. 5d).   In these data, however, sharp peaks in the LDOS exhibiting strong spatial and voltage dependence are also prominent (red features in Fig. 5d). These peaks arise due to quantum confinement of the surface state electrons within the small and regularly shaped island, similar to those of the quantum corral.\cite{Crommie1993}  Again, a clear signature  in the spectroscopy due to the bright defect is notably absent. For reference, the line profile showing the bright defect apparent height at $V_\mathrm{bias}=$1 V is overlaid on the spectra in Fig. 5d. Any modification to the LDOS due to the defect is difficult to distinguish from that of the quantum-confined Shockley electrons.

\section{Conclusion}
We have reported a detailed STM study of natively occurring defects in UHV grown graphene on Cu(111).  We focused on two types of point defects we consistently observe within otherwise pristine graphene islands and classify them as bright and dark, according to how they image at most biases, but specifically at $V_\mathrm{bias}=$1 V.  Atomic resolution images achieved with functionalized tips reveal that bright defects have a trigonal symmetry surrounding a C site in the graphene lattice, which is rotated by 60$^\circ$ depending on the sublattice assignment of the C site. These defects are identified as single vacancies arising in the as-grown graphene.  The graphene lattice is undisturbed by dark defects, suggesting that these are point defects in the underlying Cu lattice.  dI/dV spectroscopy of the defects shows their primary electronic signature is in modifying the Shockley surface state of the Cu(111) substrate.  More subtle effects of the bright lattice defects on graphene's LDOS may not be observable for this experimental system.  On the other hand, experiments on UHV grown graphene have the advantage of eliminating possible environmental effects, where contaminants could easily functionalize C vacancies.  This experiment also provides a study of graphene point defects arising from a growth procedure rather than ion bombardment or irradiation.  We expect this work to provide new perspective on the creation and imaging of point defects in graphene and motivation for pursuing growth as a means to tune defect density for controllable doping and functionalization for graphene devices.

 \begin{acknowledgements} 
Funding for this research was provided by the Center for Emergent Materials at the Ohio State University, an NSF MRSEC (Award Numbers DMR-1420451 and DMR-0820414).  
\end{acknowledgements}

\clearpage

\begin{figure}
\begin{center}
\includegraphics[width=0.8\columnwidth,keepaspectratio]{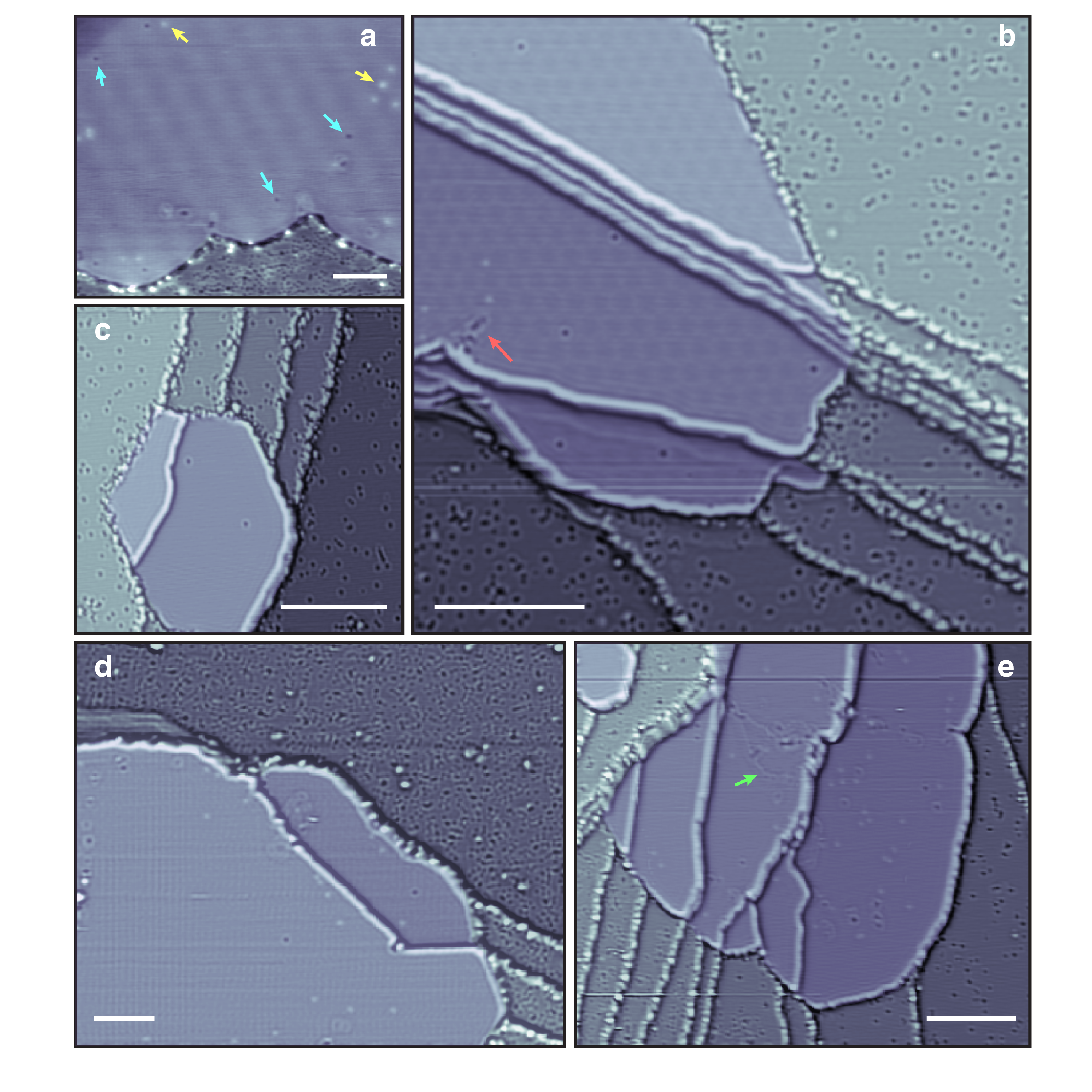}
\caption{STM images of UHV-grown graphene islands on Cu(111). Yellow arrows indicate examples of bright defects and blue arrows indicate examples of dark defects. The red arrow in (b) points to a cluster of dark defects and the green arrow in (e) points to a graphene grain boundary.  Images in (b-e) are topographic images overlaid with images treated with a Laplacian filter to bring out small differences in height.  All scale bars are 10 nm.  Images (a-d) taken at $V_\mathrm{bias}=$1 V and $I=$0.2 nA. Image (e) taken at $V_\mathrm{bias}=$1 V and $I=$1 nA.
\label{cap:fig1}}
\end{center}
\end{figure}

\begin{figure}
\begin{center}
\includegraphics[width=0.8\columnwidth,keepaspectratio]{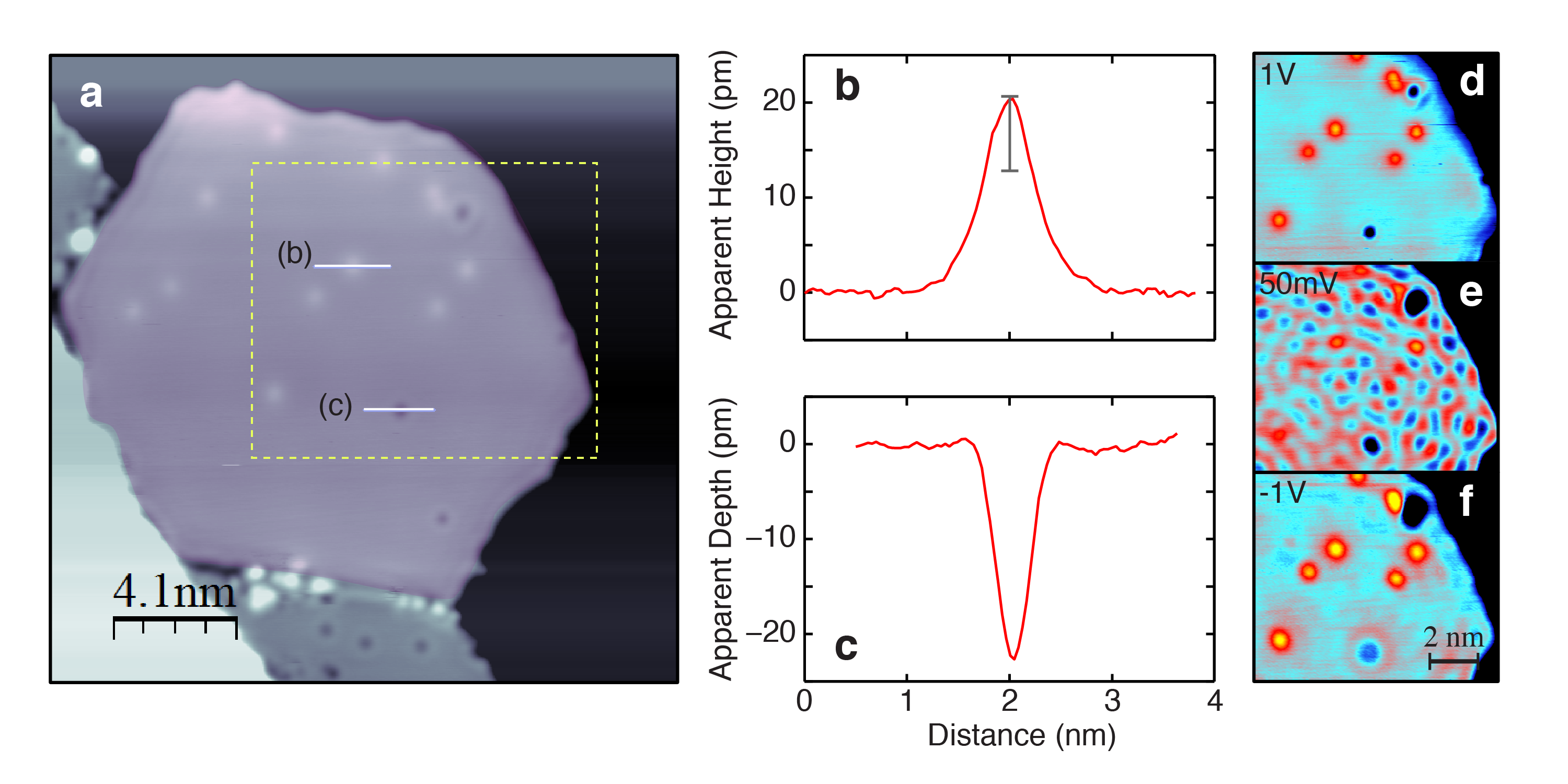}
\caption{Graphene point defects. a) STM image at 1 V, 0.2 nA of a graphene island with 11 bright defects and 3 dark defects b) and c) depict example apparent height and apparent depth profiles, respectively, of the defects marked with horizontal white lines in (a) The vertical grey bar in (b) represents the standard deviation around the mean apparent height for bright defects in (a).  d-f) Bias dependence of defect appearance for images taken at 0.2 nA and d) 1V, e) 50 mV, and f) -1V for the region outlined by the yellow box in (a). Color scale varies from 0 (black) to 36 pm (yellow).  
\label{cap:fig2}}
\end{center}
\end{figure}

\begin{figure}
\begin{center}
\includegraphics[width=0.5\columnwidth,keepaspectratio]{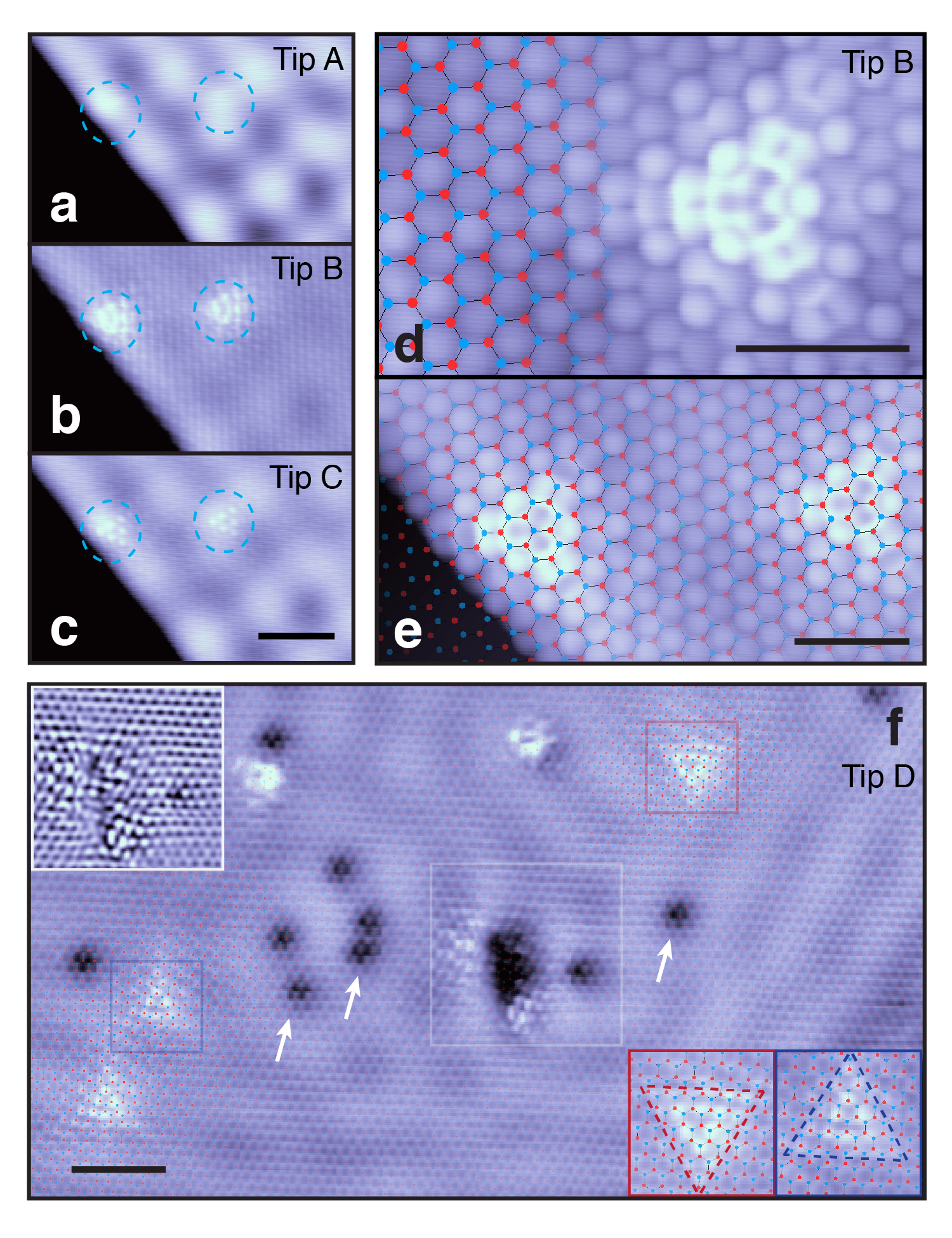}
\caption{Atomic resolution imaging of point defects. a-c) STM images of a two bright defects (blue dashed circles) imaged at $V_\mathrm{bias}=$10 mV and $I=$1 nA with a) a metal-terminated tip (Tip A), b) an unknown non-metal terminated tip (Tip B), and c) a CO terminated tip (Tip C). (Scale bar is 2 nm) d-e) Same defects as (a-c) imaged with Tip B using tunneling conditions that produce very high resolution of the defect structure and surrounding lattice. The lattice overlay shows that the defect is centered on a C site. ($V_\mathrm{bias}=$10 mV and $I=$0.2 nA, scale bar is 1 nm) f) Atomic resolution image with a non-metal terminated tip (Tip D) and lattice overlay of a large graphene region with a number of dark defects and three bright defects. Lower insets: Magnified view of the two orientations of triangular bright defects showing their corresponding sublattices (red and blue). Upper inset: Laplace filtered image showing lattice deformations in the central dark defective region. ($V_\mathrm{bias}=$106 mV and $I=$0.4 nA, scale bar is 2 nm)
\label{cap:fig3}}
\end{center}
\end{figure}

\begin{figure}
\begin{center}
\includegraphics[width=0.8\columnwidth,keepaspectratio]{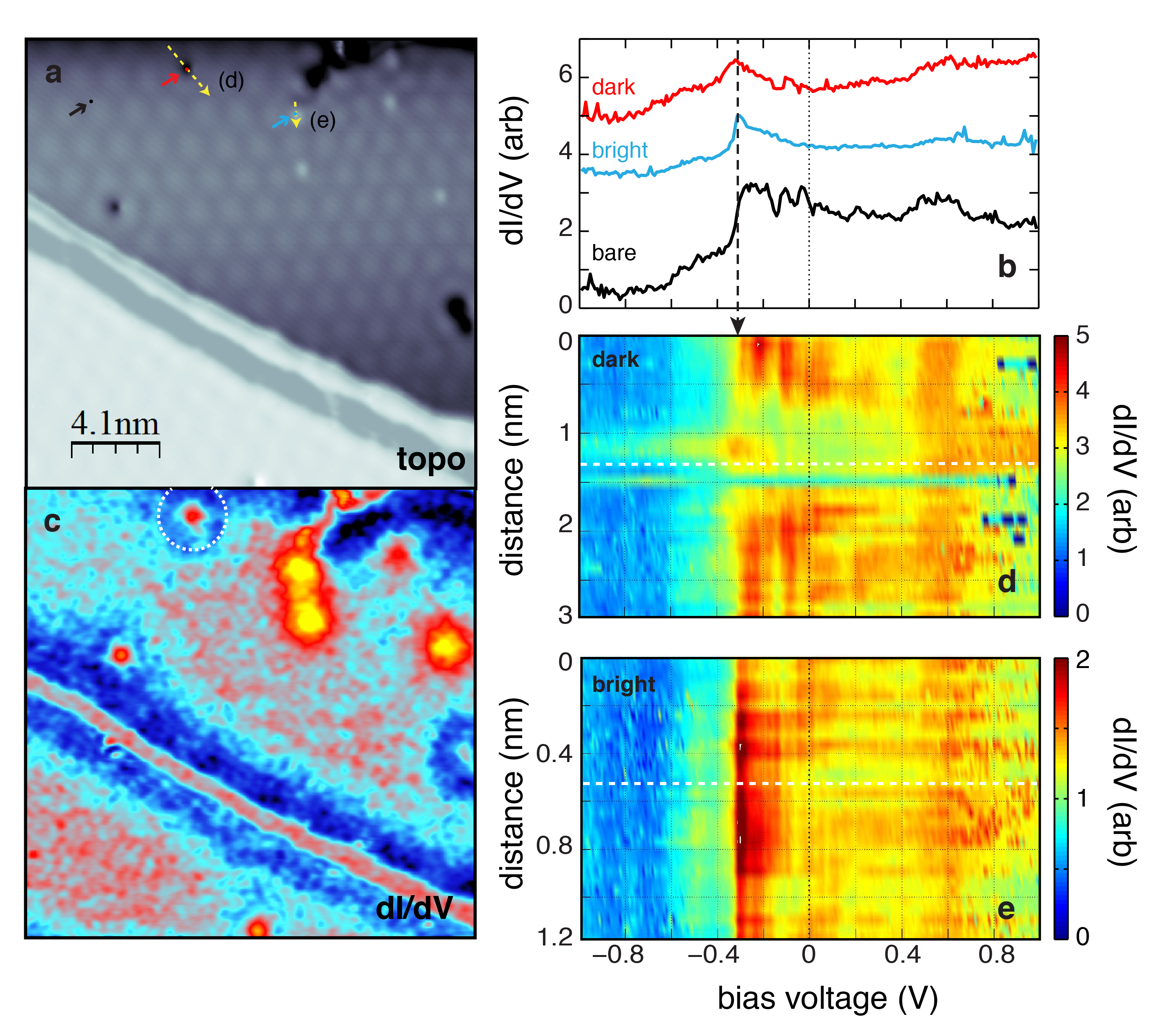}
\caption{Spectroscopy of point defects. a) Topographic image of a graphene island with 5 bright defects and 3 dark defects (one on upper terrace).  A 1.5 nm Moire pattern is visible.  ($V_\mathrm{bias}=$-0.31 V, $I=$0.5 nA) b) dI/dV point spectroscopy to measure the local density of states on dark (red) and bright (blue) defects compared to bare graphene (black), at locations marked in (a). Dashed black line at -0.31 V marks the Shockley surface state band edge. c) Spatial dependence of dI/dV of region in (a) at $V_\mathrm{bias}=$-0.31 V and $I=$0.5 nA. d-e) Spatial and voltage dependence of dI/dV in dense lines (representing 30 spectra each) over d) a dark defect and e) a bright defect in (a) (yellow dashed arrows). 
\label{cap:fig4}}
\end{center}
\end{figure}

\begin{figure}
\begin{center}
\includegraphics[width=0.5\columnwidth,keepaspectratio]{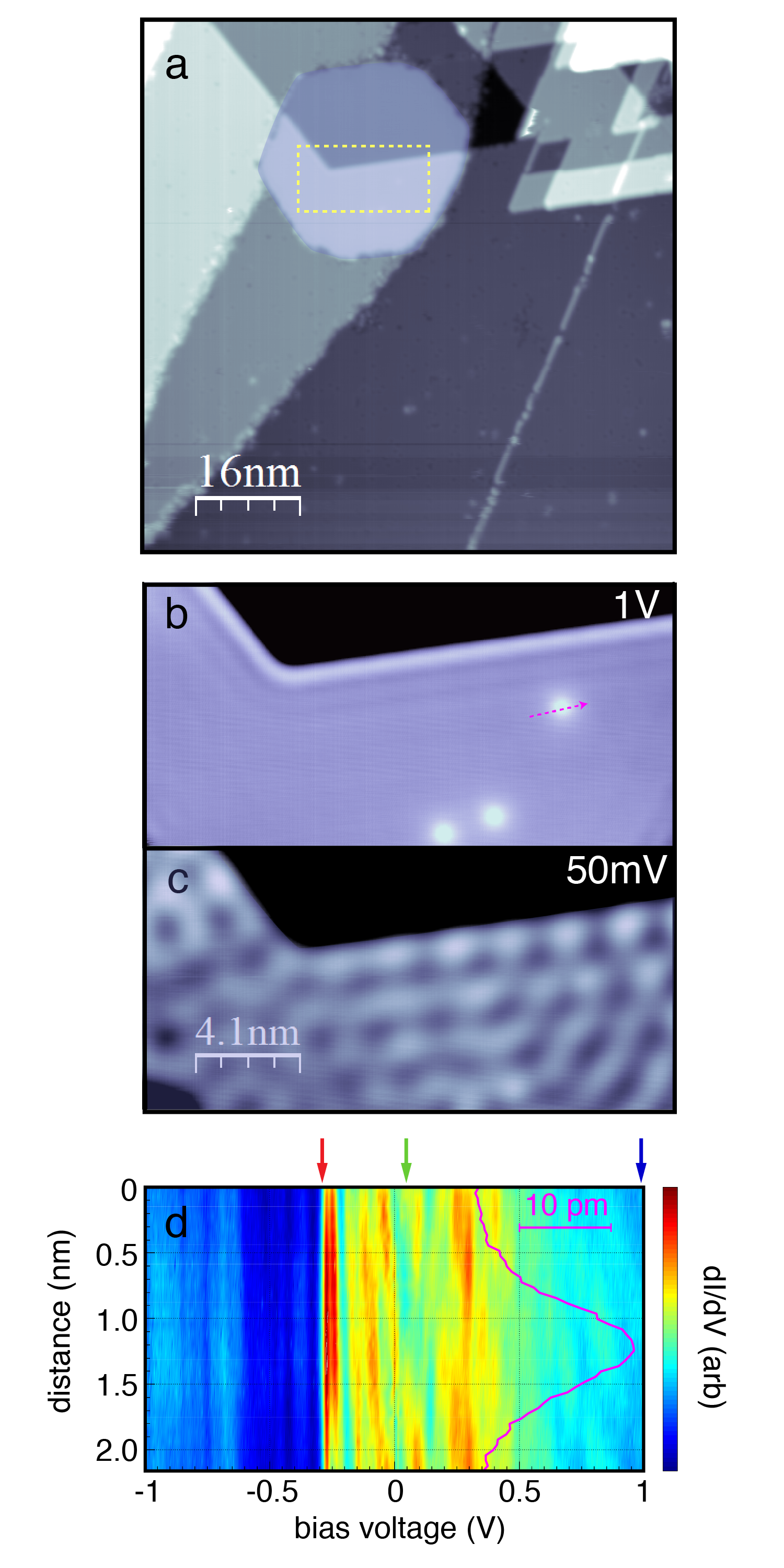}
\caption{Quantum confinement effects on spectroscopy. a) STM image of a small graphene island with regular hexagonal edges.  A Cu step edge with a 120$^\circ$ bend bisects the island. b-c) STM image showing a magnified view of a section of the upper terrace of the island (yellow dashed rectangle in (a)) with three bright defects imaged at $I=$1 nA and b) $V_\mathrm{bias}=$1 V and c) 50 mV. Standing waves due to quantum confinement of Shockley surface state electrons is visible in (c).  d) dI/dV spectroscopy in a line over a bright defect with topographic line profile at 1 V overlaid (pink dashed arrow in (b)). Quantum confinement peaks show up as spatially varying, bright peaks in dI/dV. Red arrow marks the Shockley surface state band edge, blue and green arrows mark the voltages at which images in (b) and (c) were taken, respectively. 
\label{cap:fig5}}
\end{center}
\end{figure}

\end{document}